\documentclass{article}

\usepackage[final]{nips_2017}

\usepackage[utf8]{inputenc} 
\usepackage[T1]{fontenc}    
\usepackage{hyperref}       
\usepackage{url}            
\usepackage{booktabs}       
\usepackage{amsfonts}       
\usepackage{nicefrac}       
\usepackage{microtype}      

\usepackage{multirow}
\setcitestyle{numbers}
\setcitestyle{square}
\setcitestyle{comma}
\usepackage{graphicx}

\title{Unobtrusive and Multimodal Approach for Behavioral Engagement Detection of Students}

\author{
  Nese Alyuz \\
  Intel Labs\\
  Anticipatory Computing Lab \\
  Hillsboro, OR, USA \\
  \texttt{nese.alyuz.civitci@intel.com} \\
  \And
  Eda Okur\\
  Intel Labs\\
  Anticipatory Computing Lab \\
  Hillsboro, OR, USA \\
  \texttt{eda.okur@intel.com} \\
  \And
  Utku Genc \\
  Intel Labs\\
  Anticipatory Computing Lab \\
  Hillsboro, OR, USA \\
  \texttt{utku.genc@intel.com} \\
  \And
  Sinem Aslan \\
  Intel Labs\\
  Anticipatory Computing Lab \\
  Hillsboro, OR, USA \\
  \texttt{sinem.aslan@intel.com} \\
  \And
  Cagri Tanriover \\
  Intel Labs\\
  Anticipatory Computing Lab \\
  Hillsboro, OR, USA \\
  \texttt{cagri.tanriover@intel.com} \\
  \And
  Asli Arslan Esme \\
  Intel Labs\\
  Anticipatory Computing Lab \\
  Hillsboro, OR, USA \\
  \texttt{asli.arslan.esme@intel.com} \\
}

\begin{document}

\maketitle

\begin{abstract}
  We propose a multimodal approach for detection of students’ behavioral engagement states (i.e., \textit{On-Task} vs. \textit{Off-Task}), based on three unobtrusive modalities: \textit{Appearance}, \textit{Context-Performance}, and \textit{Mouse}. Final behavioral engagement states are achieved by fusing modality-specific classifiers at the decision level. Various experiments were conducted on a student dataset collected in an authentic classroom.
\end{abstract}

\section{Introduction}

Student engagement in learning is critical to achieving positive learning outcomes \cite{Carini-2006}. Fredricks \textit{et al}. \cite{Fredricks-2004} framed student engagement in three dimensions:  Behavioral, emotional, and cognitive. In this work, we focus on behavioral engagement, where we aim to detect whether a student is \textit{On-Task} or \textit{Off-Task} \cite{Pekrun-2012, Rodrigo-2013} at any time of the learning task. Towards this end, we propose a multimodal approach for detection of students’ behavioral engagement states (i.e., \textit{On-Task} vs. \textit{Off-Task}), based on three unobtrusive modalities: \textit{Appearance}, \textit{Context-Performance}, and \textit{Mouse}. Final outputs of behavioral engagement states are obtained by fusing modality-specific classifiers at the decision level.

\section{Methodology}
\label{method}

The proposed detection scheme incorporates data collected from three unobtrusive modalities: (1) \textit{Appearance}: upper-body video captured using a camera; (2) \textit{Context-Performance}: students’ interaction and performance data related to learning content; (3) \textit{Mouse}: data related with mouse movements during the learning process. For a better evaluation of results, we analyzed the results separately for two learning tasks available: (1) \textit{Instructional}, where students are watching videos; and (2) \textit{Assessment}, where students are solving related questions.


Modality-specific data are fed into dedicated feature extractors \cite{OpenCV-2000, Pardos-2014, Mouse-2016}, and the features are then classified with respective uni-modal classifiers (i.e., Random Forest Classifiers \cite{RF-2004}). The decisions of separate classifiers are fused to output a final behavioral engagement state. For fusion, we propose to obtain a decision pool by incorporating all decision trees of modality-specific random forests and compute majority voting. This is equivalent to summing modality-specific confidence values and selecting the label with the highest confidence. Further details of the modalities, extracted features, and various fusion approaches we explored can be found in the full version of this paper \cite{MIE-ICMI-2017}.


\section{Experimental Results}
\label{exp_res}

Through authentic classroom pilots, data were collected while the students were consuming digital learning materials for Math on laptops equipped with cameras. In total, 113 hours of data were collected from 17 9\textsuperscript{th} graders for 13 sessions (40 minutes each), including the three unobtrusive modalities. For feature extraction, a sliding window of 8-seconds with 4-second overlap was utilized as in \cite{AIED-2017} for each modality. The collected data were labeled using HELP \cite{ET-2017} by three educational experts. For final ground truth labels, the windowing was also applied over three label sets, which was followed by majority voting and validity filtering. 

For the classification experiments, we divided each student’s data into 80\% and 20\% partitions, for training and testing, respectively. In order to reduce the effect of overfitting, we conducted leave-one-subject-out cross-validation and applied 10-fold random selection to balance training sets. The uni-modal and fusion results for different learning tasks (averaged over all runs and all student) are summarized in Table~\ref{T1}. As these results indicate, for \textit{Instructional} sections, \textit{Appearance} modality performs best (0.74) due to the lack of interactions necessary for the other modalities. For \textit{Assessment}, fusing all modalities yields the best performance (0.89).

\begin{table}[!h]
  \caption{F1-measures for uni-modal models (Appr: \textit{Appearance}, CP: \textit{Context-Performance}, Ms: \textit{Mouse}) and fusion (INSTR.: \textit{Instructional}, ASSESS.: \textit{Assessment}).}
  \label{T1}
  \centering
  \begin{tabular}{*6c}
    \toprule
    \textbf{Section Type} & \textbf{Class} & \textbf{Appr} & \textbf{CP} & \textbf{Ms} & \textbf{FUSION} \\
    \toprule
    \textbf{INSTR.} & \textbf{On-Task} & \textbf{0.78} & 0.66 & 0.60 & 0.75 \\
     & \textbf{Off-Task} & \textbf{0.71} & 0.50 & 0.51 & 0.64 \\
    \cmidrule(r){2-2}\cmidrule(lr){3-5}\cmidrule(l){6-6}
     & \textbf{OVERALL} & \textbf{0.74} & 0.59 & 0.54 & 0.70 \\
    \midrule
    \textbf{ASSESS.} & \textbf{On-Task} & 0.87 & 0.86 & 0.87 & \textbf{0.93} \\
     & \textbf{Off-Task} & 0.66 & 0.22 & 0.64 & \textbf{0.73} \\
    \cmidrule(r){2-2}\cmidrule(lr){3-5}\cmidrule(l){6-6}
     & \textbf{OVERALL} & 0.81 & 0.74 & 0.80 & \textbf{0.89} \\
    \bottomrule
  \end{tabular}
\end{table}

\section{Conclusion}

In summary, for behavioral engagement, we get relatively high results when only \textit{Appearance} modality is used for \textit{Instructional} sections whereas the fusion of all modalities yields better results in \textit{Assessment} sections. The experiments also showed that it is beneficial to have context-dependent classification pipelines for different section types (i.e., \textit{Instructional} and \textit{Assessment}). In the light of these results, we can say that context plays an important role even when different tasks in the same vertical (i.e., learning) are considered.

\small

\bibliography{main}

\end{document}